# Magnetic Field Driven Quantum Phase Transitions in Josephson Arrays

J. Paramanandam, M.T. Bell, L.B. Ioffe, and M.E. Gershenson

Department of Physics and Astronomy, Rutgers University,

Piscataway, New Jersey 08854

Abstract

We have studied the magnetic-field-driven quantum phase transitions in Josephson junction arrays with a large coordination number. The characteristic energies were extracted in both the superconducting and insulating phases by integrating the current-voltage characteristics over a voltage range $2eV \leq k_B T$. For the arrays with a relatively strong Josephson coupling, we observed duality between the energies in the superconducting and insulating phases. The arrays with a weaker Josephson coupling demonstrate an intermediate, "bad metal" regime in weak magnetic fields; this observation underlines the importance of vortex pinning at large scales and, presumably, emergent inhomogeneity in the presence of strong offset charge disorder.

The arrays of small superconducting islands coupled by Josephson junctions (referred to below as Josephson arrays) represent a quintessential system for the study of the "bosonic" superconductor-insulator transition (SIT) that is driven by the enhancement of phase fluctuations induced by Coulomb interactions [1,2,3,4]. The essence of this SIT is the localization of Cooper pairs, which occurs when the coupling between superconducting islands (characterized by the Josephson energy $E_J$) becomes weaker than the on-site repulsion of Cooper pairs (characterized by the Coulomb energy $E_C$). After more than three decades of research, this phenomenon remains poorly understood, which reflects the complexity of quantum phase transitions in strongly disordered interacting systems. The ubiquitous disorder in Josephson arrays is associated with random (and fluctuating) offset charges on superconducting islands (see, e.g., Ref. 5).

Prior experiments with Josephson arrays [1,6,7] made a puzzling observation: when the Josephson coupling between islands was frustrated by a (weak) magnetic field $B$, the superconducting phase was not directly transformed into the insulating one. Instead, an intermediate, "bad metal" regime was observed over a relatively wide range of fields: the resistance $R$ in this regime increased with $B$, but, at the same time, remained almost $T$-independent. This is in contrast with the "direct" magnetic-field-driven SIT observed for thin disordered films [8,9,10,11,12].

In this Letter, we address this long-standing problem using a new tool: Josephson arrays with a large number of interacting nearest-neighbor islands ($N = 10$, see the inset in Fig. 2c). Theoretically, one expects that the behavior of these arrays is simpler because it is described by the mean field theory. On the experimental side, these arrays compare favorably with the conventional ($N = 4$) arrays due to a better averaging of the junction parameters. Our measurements were focused on the Cooper pair transport at sufficiently low temperatures ($T \leq 0.2$ K) such that the density of equilibrium quasiparticles on superconducting islands becomes negligible. Given a limited $T$ interval available for the analysis of the $R(T)$ dependences, we have determined the characteristic energy scale, $T_0$, by integrating the current-voltage characteristics (IVCs) over a voltage range $2eV \leq k_B T$. This novel method enables finding $T_0$ even if this energy is smaller than the physical temperature. For the arrays with stronger Josephson coupling, we observed for the first time the *direct* magnetic-field-driven SIT, reminisent of that in thin disordered films and in line with the "dirty boson" scenario developed for weakly-disordered arrays [3,4,13]. For the arrays with weaker Josephson coupling, and, thus, weaker critical magnetic fields, an intermediate "bad metal" regime was observed. This observation implies the appearance of new physics at large scales, which, presumably, is due to vortex pinning by emergent inhomogeneity in the presence of strong offset charge disorder.

The arrays were fabricated by the multi-angle electron-gun deposition of aluminum films through a lift-off mask (for details, see Refs. 14,15). The arrays consisted of two sets of overlapping superconducting wires ("islands"), with Josephson junctions formed at each intersection between the wires; each island overlapped with 10 other islands ($N = 10$). The larger-scale structure of the arrays included 8×8 "super-cells" shown in the inset of Fig. 2c; each super-cell is connected to its neighbors with four (five) common vertical (horizontal) wires. The current leads were attached in the bus-bar geometry to the opposite sides of the arrays. The uniformity of the junction parameters across the whole array was verified by observing a very "sharp" magnetic-field dependence of the critical super-current $I_C$ for arrays in the superconducting state: $I_C$ diminishes by approximately 50% at $B$ that corresponded to one flux quantum per the whole array (see the inset of Fig. 1b).

Below we focus on three representative arrays (A, B, and C) with different Josephson coupling between the islands. Parameters of these arrays are listed in the Table. The charging energy $E_C^j \equiv \frac{e^2}{2C_J}$ for individual junctions was estimated using the junction area and the specific capacitance for Al tunnel junctions (50 fF/μm$^2$, see e.g., Ref. 6). The Josephson energy per junction, $E_J^j$, was calculated on the basis of the Ambegaokar-Baratoff relationship [16], the junction normal-state resistance, and the critical temperature of Al film, ~1.3K. The competition between the Josephson coupling and charging effects in the arrays is controlled by the parameter $\frac{E_J}{E_C}(B=0) \equiv N^2 \frac{E_J^j}{E_C^j}$. Here $E_J = NE_J^j$ is the effective Josephson energy of $N$ Josephson junctions connected to a single superconducting island, and $E_C = E_C^j/N$ is the effective charging energy per island.

The arrays were studied in a μ-metal-shielded dilution refrigerator; a weak magnetic field was applied perpendicular to the array's plane using a superconducting solenoid. The current-voltage characteristics were measured in the four-probe configuration using a *dc* current source and an amplifier with a large (1GΩ) input resistance. The differential resistance $\frac{dV}{dI}$ was obtained using the lock-in technique at frequencies of 6-13 Hz. The arrays were shielded from *rf* noises and microwave radiation by cold filters.

| Array | unit cell area, $a^2$, µm² | Junction resistance $R_j(2K)$, kΩ | Array resistance $R(2K)$, kΩ | $E_J = NE_J^j$, K | $E_C = \frac{E_C^j}{N}$, K | $\frac{E_J}{E_C}(B=0) \equiv N^2 \frac{E_J^j}{E_C^j}$ |
|---|---|---|---|---|---|---|
| A | 1 | 43.5 | 5.0 | 1.7 | 0.092 | 18.5 |
| B | 1 | 73 | 8.4 | 1.0 | 0.14 | 7.1 |
| C | 2 | 133 | 15.2 | 0.55 | 0.13 | 4.2 |

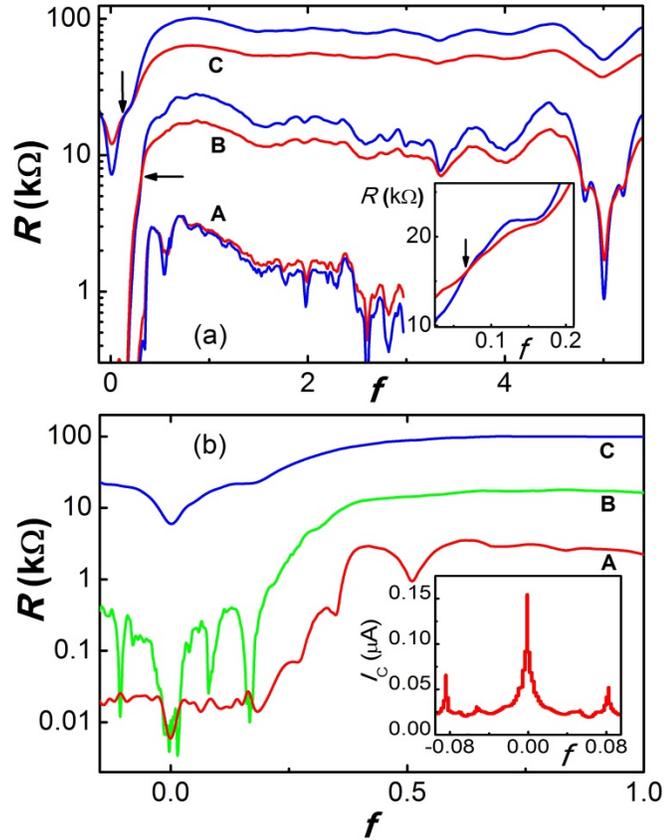

**Fig. 1**. *Panel a*: The magnetic field dependences of the zero-bias resistance measured for arrays A, B, and C at $T = 50$mK (blue curves) and 100mK (red curves). The magnetic field $f = B/B^*$ is normalized by $B^* = \Phi_0/(10a^2)$ (see the text). In particular, $f = 1$ corresponds to $B \cong 2$ G for arrays A and B, and $B \cong 1$ G for array C. The inset in panel *a* shows the blow-up of the $R(f)$ dependence for array C at small $f$. The low-field crossings between the $R(T)$ dependences measured at different $T$ are marked by arrows.
*Panel b*: The $R(f)$ dependences for arrays A - C at $T = 50$mK in weak magnetic fields. The inset in panel *b* shows the magnetic field dependence of the critical current (at which the voltage drop across the sample exceeds 0.5 µV) for array A at $T = 50$mK.

The "zero-bias" resistance (obtained at small bias currents of 3-50 pA corresponding to the Ohmic regime), measured as a function of magnetic field at $T = 50$ mK and 100 mK, is shown for all arrays in Fig. 1a. Below we use the normalized values of the magnetic field $f = B/B^*$, where $B^* = \Phi_0/(Na^2)$, $a$ is the distance between the adjacent parallel wires, $\Phi_0 \equiv h/2e$ is the flux quantum. The $R(f)$ dependences are periodic with the period $f = N$ (i.e. $\Phi = \Phi_0$ for a unit cell); the local minima of $R(f)$ occur at rational values of $f$ when the vortex lattice period is commensurate to the underlying array structure. Each crossing between the $R(f)$ dependences measured at different $T$ corresponds to a transition between the "insulating" and "superconducting" phases at a critical magnetic field $f_C$ (for clarity, only two $R(f)$ dependences are shown). Several transitions with different values of the critical resistance $R_C$ have also been observed in experiments with conventional ($N = 4$) arrays near resistance minima at rational $f$ values (see e.g., Ref. 7). The $R(T)$ dependences for arrays A-C are shown in Fig. 2. We do not observe saturation of $R(T)$ down to lowest temperatures; this is in contrast with the low-$T$ saturation of $R(T)$ observed in experiments [6,7] and attributed to finite-size effects. Whereas array A with a larger $E_J$ remains on the borderline between superconducting and insulating phases even at the maximum frustration, arrays B and C enter the insulating phase with increasing $B$.

The $R(T, f = \text{const})$ analysis is insufficient for the detailed study of the observed transitions because of the limited $T$ range. Indeed, the "freeze-out" of equilibrium quasiparticles in the islands occurs only below 0.2 K, and quantum fluctuations that drive the SIT become significant at approximately the same $T$ (the values of $k_B T \approx E_C$ for our arrays are shown in Fig. 2 by arrows). More information can be obtained from the analysis of the current-voltage characteristics at the base temperature. The differential resistance $\frac{dV}{dI}$ measured as a function of the bias current $I$ at several $f$ values near the critical field $f_C$ is shown in Fig. 3. Depending on the sign of the curvature $\frac{d^2V}{dI^2}(I = 0)$, all $\frac{dV}{dI}(I)$ dependences can be sorted into two categories. The dependences with $\frac{d^2V}{dI^2}(I = 0) > 0$ correspond to the "superconducting" phase. This IVC non-linearity is the precursor of the super-current that develops on the superconducting side of the SIT. The negative curvature $\frac{d^2V}{dI^2}(I = 0)$ is a signature of the insulating phase characterized by the IVCs with a Coulomb-blockade structure. Below we analyze the integrals (1) and (2), which also characterize the IVC nonlinearity and provide direct measure of energy scales.

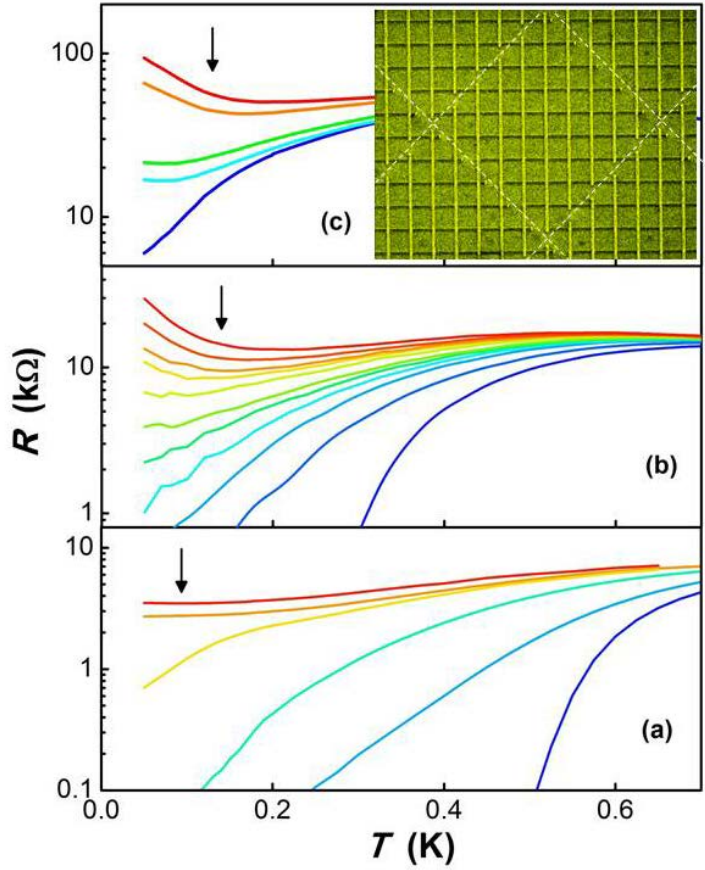

**Fig. 2.** Temperature dependences of the zero-bias resistance for arrays A, B, and C (panels *a*, *b*, and *c*, respectively) measured at several fixed values of magnetic fields varied from *f* = 0 (blue curves) to the field corresponding to the maximum resistance for arrays B and C, and the first maximum of *R*(*f*) for array A (Fig. 1a). Temperatures $T = \frac{E_C}{k_B}$ are shown by arrows. The inset in panel *c* shows the micrograph of the diamond-shaped super-cell of the arrays. Qualitatively, these arrays can be viewed as the network of wire "bundles" in which order parameter is associated with individual wires and each wire interacts will all others in a crossing bundle.

In order to extract the characteristic energy $T_0$ from the IVCs, we focus on the low bias currents $|I| \leq I^*$ which correspond to the voltage drop across the array $\leq k_B T/2e$ (the shape of dependences $\frac{dV}{dI}(I)$ at $|I| > I^*$ is approximately the same in both the insulating and superconducting/bad metal phases). By subtracting $\frac{dV}{dI}(I = I^*)$ from the corresponding $\frac{dV}{dI}(I)$ dependences, we obtained the set of dependences $\frac{dV}{dI}(I) - \frac{dV}{dI}(I = I^*)$ intersecting at $I = I^*$. For all "insulating" IVCs with negative curvature at $I = 0$, we define the $T_0$ values as follows:

$$T_0(f) = \frac{2e}{k_B} \times \int_0^{I^*} \left\{ \frac{dV}{dI}(f, I) - \frac{dV}{dI}(f, I^*) \right\} dI . \qquad (1)$$

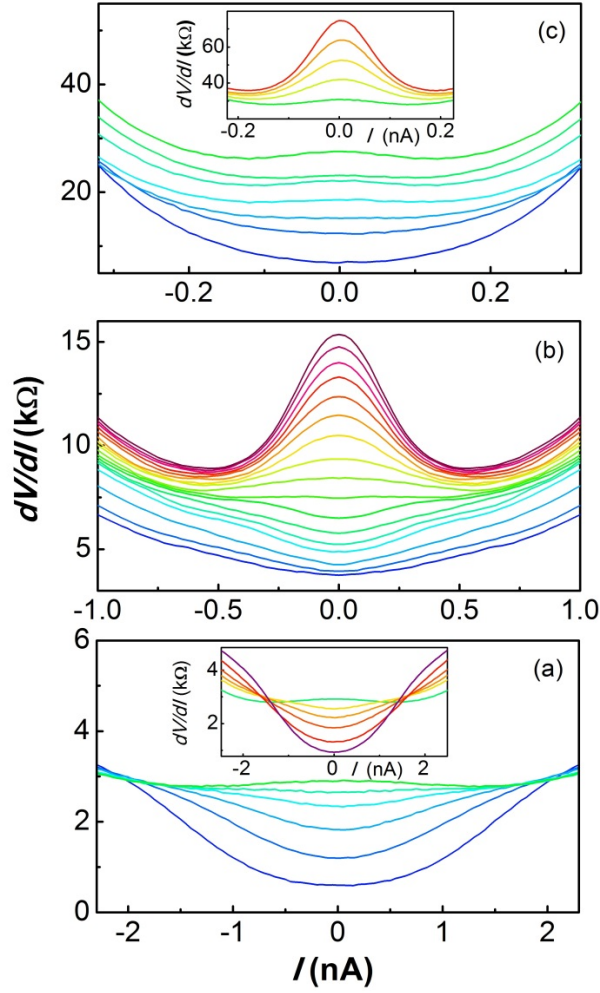

**Fig. 3.** Dependences $\frac{dV}{dI}(I)$ measured for arrays A-C (panels *a*, *b*, *c*, respectively) at $T = 50$ mK and several values of magnetic field close to $f_C$ (the weakest magnetic field corresponds to blue curves, the strongest – to red curves, the field increments are the same for each panel). Because the dependence $\frac{dV}{dI}(I = 0, f)$ is non-monotonic for array A, the main panel of Fig. 3a shows the data up to $f$ corresponding to the maximum of $\frac{dV}{dI}(I) = 0$, and the inset – the data for stronger fields. This non-monotonicity is less pronounced for array C (the main panel of Fig. 3c); the inset in Fig. 3c shows the data at stronger fields in the insulating phase.

Deeply in the insulating phase where the "Coulomb" gap is fully developed ($T_0 \gg T$), this procedure is equivalent to the extraction of $T_0$ from the "offset" voltage $(T_0 = 2eV_{offset}/k_B)$ in the Coulomb blockade regime [17,18].

In order to extract the relevant energy scale in the superconducting regime, we define the dual quantity:

$$T_0(f) = \frac{2eR_C}{k_B} \times \int_0^{R_C I^*} \left\{ \frac{dI}{dV}(f,I) - \frac{dI}{dV}(f,I^*) \right\} dV . \quad (2)$$

Again, deeply in the superconducting phase, this energy is related to the critical current $I_C$: $T_0 = \frac{\hbar I_C}{2ek_B}$. The superconducting and insulating phases are separated by the "critical" IVC with $\frac{d^2V}{dI^2}(I=0) \approx 0$ at $f=f_C$. Note that the selection of the critical IVC does not affect the shape of $T_0(f)$ dependence, though it may result in a small vertical shift of the curves shown in Fig. 4. With increasing temperature, the non-linearity of IVCs at small biases is reduced, and $T_0$ is suppressed (see the top inset in Fig. 4).

The $T_0(f)$ dependences for arrays A-C are shown in Fig. 4. It is instructive to compare these data with the $R(f)$ dependences in Fig. 1, for both types of dependences are strongly affected by the commensurability effects. Vortex pinning at rational $f$ stabilizes the superconducting phase. For example, strong pinning effects (clearly seen at $f = 0.5$) prevent array A from entering the insulating phase. The arrays with a smaller ratio $\frac{E_J}{E_C}(f=0)$ are driven into the insulating phase by weaker magnetic fields; this translates into a larger length scale for pinning at $f$ close to $f_C$. For example, a local $T_0(f)$ minimum observed for array C at $f \approx 0.16$ (see the bottom inset in Fig. 4) corresponds to one flux quantum per one super-cell. Local $R(T)$ minima for array B are clearly seen at $f \approx 0.16$ and $0.08$ in Fig. 1b.

The key result of the present work is the $T_0(f)$ dependences of arrays demonstrating the SIT. A single SIT is observed for array B at $f_C$ which is relatively far from all "commensurate" $f$ values. To our knowledge, this is the first observation of the direct SIT in Josephson arrays frustrated by the magnetic field. In contrast, array C remains in an intermediate "bad metal" regime over a rather extended field range. In this regime, the resistance is almost $T$-independent ($T_0 \approx 0$), although it increases with magnetic field by a factor of ~ 1.5. Beyond this field range, the $T_0$ values are rapidly growing in both "superconducting" and "insulating" phases.

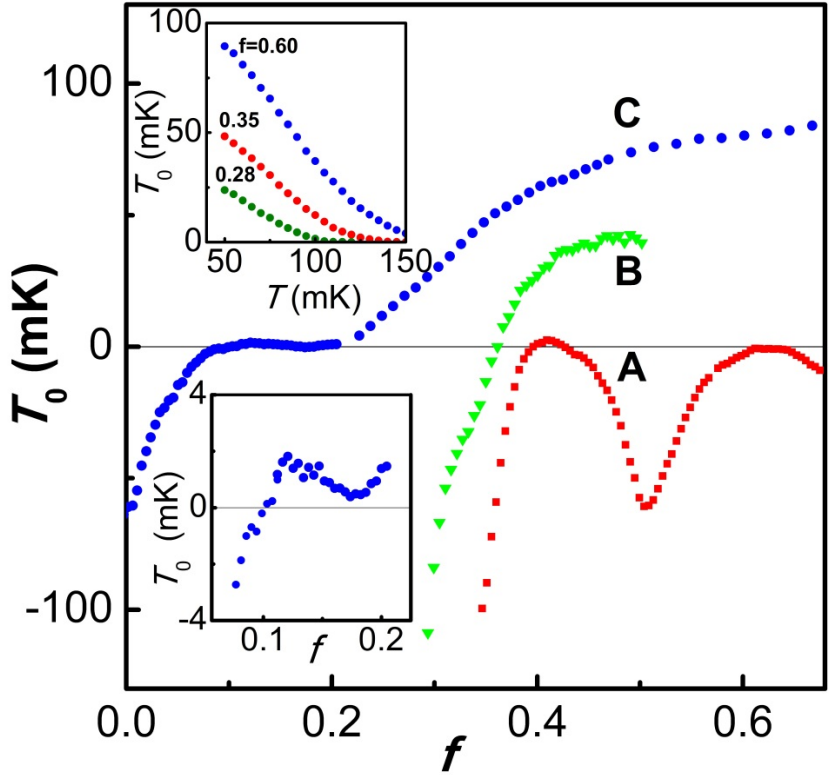

**Fig. 4.** The characteristic energy $T_0$ extracted from the IVCs measured at different $f$ and $T = 50$ mK for arrays A (■), B (▼), and C (●). The bottom inset shows the blow-up of the $T_0(f)$ dependence for array C. The top inset shows the $T_0(T)$ dependences for array C at three $f$ values corresponding to the insulating phase.

Understanding of the "bad metal" regime, observed in prior experiments with conventional arrays and in the present experiments with weakly coupled large-$N$ arrays, remains a challenge. As it was discussed above, the commensurability effects play an important role in shaping the $T_0(B)$ dependences. Note, however, that within the magnetic field range corresponding to the "bad metal" regime in array C, only two "commensurate" values of $f$ (0.16 and 0.08, corresponding to one flux quantum per one and two super-cells, respectively) should affect $T_0$ similar to $R(f)$ for array B in Fig. 1b. Between these two values of $f$, an emerging insulating phase was expected, in contrast with our experimental findings. An enhanced vortex pinning, not associated with the array structure, can be caused by a large-scale inhomogeneity which may emerge near the SIT due to random offset charges [5]. Spontaneous formation of large scale structures in nominally uniform disordered systems appears in a variety of mathematical models such as the disorder-driven SIT in films [19,20]. This emergent inhomogeneity would promote the "bad metal" regime over a wider range of weak magnetic fields. More experiments are required to test this scenario.

In conclusion, we have observed the magnetic-field-driven phase transitions in Josephson arrays with a large number of nearest-neighbor islands. The arrays with relatively strong Josephson coupling demonstrate symmetry between the characteristic energies $T_0$ in the superconducting and insulating states across the SIT. This duality, however, is lacking for weakly coupled arrays, where an intermediate "bad metal" regime with negligibly small $T_0$ is observed over a sizable range of magnetic fields. The fact that this phase appears only in weak magnetic fields underlines the importance of large-scale commensurability effects and, presumably, emergent inhomogeneity near the SIT.

We would like to thank M. Feigel'man, S. Korshunov, and B. Spivak for helpful discussions. The work was supported by the NSF (NIRT ECS-0608842 and DMR-1006265), DARPA (HR0011-09-1-0009), and ARO (W911NF-09-1-0395).